\documentclass[reprint,aps,pre,amsmath,amssymb,superscriptaddress,nofootinbib,longbibliography]{revtex4-2}

\usepackage{graphicx}
\graphicspath{{./figs/}}
\usepackage{amsmath,amssymb,mathtools}
\usepackage{bm}
\usepackage{booktabs}
\usepackage{xcolor}
\usepackage[colorlinks=true,citecolor=blue,linkcolor=blue,urlcolor=blue]{hyperref}

\newcommand{\Pw}{\hat P_{\mathcal W}}
\newcommand{\Expec}{\mathbb{E}}
\newcommand{\sigCOE}{\sigma_{\mathrm{RMT}}}
\newcommand{\lD}{\lambda_{\Delta}}
\newcommand{\lB}{\lambda_{\mathcal W}}
\newcommand{\Tr}{\operatorname{Tr}}
\newcommand{\Var}{\operatorname{Var}}
\newtheorem{theorem}{Theorem}

\begin{document}

\title{Edge resonances and random-matrix saturation of \\global-symmetry violation in a chaotic many-body system}

\author{Ayanendu Dutta}
\email{ayanendudutta@gmail.com}
\affiliation{Department of Physics, Jadavpur University, Kolkata-700032, India}
\affiliation{Department of Physics, Presidency University, Kolkata-700073, India}

\date{\today}

\begin{abstract}
Chaotic dynamics is expected to violate any global symmetry that is not exact, and in black-hole-like systems the symmetry-violating amplitude is commonly estimated as a random sum over $e^{S}$ microstates, of size $e^{-S/2}$. We examine what this estimate does and does not capture in a controlled setting: a random-field Heisenberg chain with an exact $U(1)$ charge, broken by a transverse field of strength $\lambda$. The observable is the microcanonical amplitude $G=\langle A|\hat P_{\mathcal W}|B\rangle$ between channels of different charge, with $\hat P_{\mathcal W}$ the projector onto an energy window. We prove an exact edge representation: $G$ is a sum over eigenstate pairs straddling the window edges, weighted by inverse level differences. Together with a divided-difference expansion of the projector, it implies exact superselection at $\lambda=0$, a selection rule $|G|_{\rm typ}\propto\lambda^{\Delta Q}$, and a variance controlled by cross-sector resonances at the edges whose scaling depends on how the window is specified: for $\Delta Q=1$, $\sigma_G^2\propto\lambda$ for a window holding a fixed number of eigenstates but $\propto\lambda^2\ln(1/\lambda)$ for a fixed energy window. A resonance theory built only from $\lambda=0$ data reproduces exact diagonalization with no free parameters. Beyond the resonant regime the normalized variance becomes essentially independent of the Hilbert-space dimension $D$, and it reaches the exact finite-$D$ random-matrix value $f(1-f)D/[(D-1)(D+2)]\simeq f(1-f)\,e^{-S}$ only when the breaking-induced decay rate becomes comparable to the window width, at breaking strengths of the order of the exchange coupling, far beyond the scale $\lambda\propto D^{-1/2}$ at which the level statistics already become Wigner--Dyson.
\end{abstract}

\maketitle

% ======================================================================
\section{Introduction}
\label{sec:intro}

Global symmetries hold a special place in the physics of chaotic quantum systems. A Hamiltonian that commutes with a charge $\hat Q$ splits into independent blocks, and its chaotic properties---random-matrix level statistics~\cite{Bohigas:1984,Brody:1981,Guhr:1998}, eigenstate thermalization~\cite{Deutsch:1991,Srednicki:1994,Rigol:2008,DAlessio:2016} and its refinements for correlations among matrix elements~\cite{Foini:2019,Pappalardi:2022} and for symmetric Hamiltonians~\cite{Murthy:2023}, operator spreading and scrambling~\cite{Khemani:2018,Rakovszky:2018,Friedman:2019}---are defined sector by sector. When the symmetry is weakly broken the blocks are coupled, and a basic question is \emph{how}, and \emph{how strongly}, the chaotic eigenstates then connect states of different charge.

The question has a long history in nuclear physics, where isospin is broken by the Coulomb interaction and parity by the weak interaction. Rosenzweig and Porter introduced random matrices that interpolate between independent symmetry sectors and a single ensemble~\cite{Rosenzweig:1960}; the statistical theory of isospin mixing in compound nuclei~\cite{Harney:1986,Guhr:1990} and of fluctuations in the presence of approximate symmetries~\cite{French:1988} followed, and the large, statistically distributed parity violation observed in neutron resonances---enhanced by the small spacing of compound levels of opposite parity~\cite{Sushkov:1982}---is the textbook case of a weak symmetry-breaking interaction acting in a chaotic system~\cite{Mitchell:1999,Weidenmuller:2009}. Coupled microwave billiards~\cite{Alt:1998,Dietz:2006} and elastic resonators~\cite{Ellegaard:1996} have since realized the gradual breaking of a quantum number under laboratory control. Much of this literature characterizes the crossover through spectral statistics, which reach those of a single ensemble once the rms coupling becomes comparable to the mean level spacing~\cite{Guhr:1990,Weidenmuller:2009}, and through the spreading width $\Gamma^{\downarrow}=2\pi\overline{v^2}/d$ given by Fermi's golden rule~\cite{Harney:1986,Mitchell:1999}. Generalized Rosenzweig--Porter ensembles later sharpened the picture into three regimes---localized, non-ergodic extended, and ergodic---separated precisely by the comparison of $\Gamma^\downarrow$ with the level spacing and with the bandwidth~\cite{Kravtsov:2015,DeTomasi:2019}. In many-body lattice models, the onset of Wigner--Dyson statistics under integrability- or symmetry-breaking perturbations occurs at strengths that decrease with system size~\cite{Rabson:2004,Santos:2010,LeBlond:2021}, while the slow relaxation of the weakly broken charge is governed by a golden-rule rate $\propto\lambda^2$~\cite{Mallayya:2019,Mallayya:2021}.

A second motivation comes from gravity. Quantum gravity is believed to admit no exact global symmetries~\cite{Banks:2011,Harlow:2019PRL,Harlow:2021CMP}. Black holes are the classic argument: a black hole formed from matter of a given global charge carries no corresponding hair and evaporates thermally, so the charge cannot be conserved~\cite{Zeldovich:1976,Hawking:1975}, and Euclidean wormholes generate symmetry-violating effective interactions~\cite{Giddings:1988,Abbott:1989,Coleman:1990,Kallosh:1995}. Recent work quantifies the violation through relative entropies and replica wormholes~\cite{Chen:2021,Hsin:2021,Harlow:2021Shaghoulian}, connects it to a charged version of eigenstate thermalization~\cite{Belin:2022}, and estimates the probability of an exclusive charge-violating process mediated by a black hole of entropy $S$ as $|\mathcal A|^2\sim e^{-S}$, interpreting the amplitude as a random variable with a fluctuating phase and leaving the prefactor undetermined~\cite{Bah:2023}. Heuristically, the violating amplitude is a random walk over $\sim e^{S}$ microstate contributions of size $e^{-S}$, giving $e^{-S/2}$~\cite{Zhiboedov:notes}; the same entropy suppression of exclusive amplitudes underlies the black-hole ansatz for high-energy gravitational scattering~\cite{Giddings:2008}, and chaotic dynamics of the black-hole $S$-matrix has been argued for on general grounds~\cite{Polchinski:2015}. A random-matrix description of black-hole microstates~\cite{Cotler:2017,Saad:2019,Marolf:2020} makes this estimate natural, but it also makes clear that the estimate is a statement about eigenvector statistics: it presupposes that the microstates have already randomized across the charge sectors, and it is silent on how that randomization is approached and on the statistics of the amplitude along the way.

A third motivation is practical. Symmetry violation is a central error mode in quantum simulators of constrained dynamics---in lattice gauge theories, for instance, small gauge-breaking errors produce long-lived prethermal plateaus whose violation grows only perturbatively~\cite{Halimeh:2020}---and conservation laws are routinely used as error witnesses and for symmetry verification~\cite{BonetMonroig:2018,McArdle:2019}. Quantum processors now simulate Heisenberg chains with conserved magnetization at scale~\cite{Rosenberg:2024}, and scrambling and wormhole-inspired protocols have been realized on trapped-ion and superconducting hardware~\cite{Landsman:2019,Mi:2021,Jafferis:2022,Brown:2023}. For such experiments it matters which statistical features of symmetry violation follow from chaos alone, and which window of breaking strengths they require.

In this work we study a single, sharply posed problem that sits at the intersection of these three lines: the statistics of the microcanonical charge-violating amplitude
\begin{equation}
G\equiv\langle A|\Pw|B\rangle ,
\label{eq:G}
\end{equation}
where $|A\rangle$ and $|B\rangle$ are channels with charges $N_A\neq N_B$ and $\Pw$ projects onto the eigenstates of the full Hamiltonian $H(\lambda)$ in an energy window $\mathcal W$. Equation~\eqref{eq:G} is the quantity whose magnitude the random-walk estimate addresses: if the eigenvectors in $\mathcal W$ were Haar random in the full Hilbert space of dimension $D=e^{S}$, $G$ would be a sum of $fD$ terms of random sign and size $1/D$ ($f$ is the window fraction), so $|G|\sim\sqrt{f}\,e^{-S/2}$. We ask how $G$ behaves as a function of the breaking strength $\lambda$, from exact conservation to complete mixing, in a concrete chaotic spin chain whose $\lambda=0$ dynamics is ergodic within each charge sector. We do not model gravity; $S\equiv\ln D$ is the logarithm of a Hilbert-space dimension, and the results concern chaotic many-body systems with a weakly broken $U(1)$ symmetry. Our main results are:

(i)~An \emph{exact edge representation} (Theorem~\ref{thm:edge}): for any $\lambda$, $G$ is a sum over pairs of eigenstates of $H(\lambda)$ that straddle the edges of $\mathcal W$, weighted by the inverse of their energy difference. Symmetry violation is generated only at the window edges.

(ii)~\emph{Selection rule and edge resonances.} The typical amplitude scales as $\lambda^{\Delta Q}$, $\Delta Q=|N_A-N_B|$. The variance, by contrast, is dominated by rare near-degenerate pairs of opposite charge straddling an edge, which are not repelled at $\lambda=0$. Its scaling depends on how the window is specified: for $\Delta Q=1$, $\sigma_G^2\propto\lambda$ if $\mathcal W$ contains a fixed number of eigenstates, but $\sigma_G^2\propto\lambda^2\ln(1/\lambda)$ if $\mathcal W$ is a fixed energy interval. The difference is a size-bias (inspection-paradox) effect in how an edge selects level gaps. A resonance theory built only from $\lambda=0$ spectra and matrix elements reproduces exact diagonalization without fitting.

(iii)~A \emph{size-independent crossover}. Once the golden-rule width $\Gamma\propto\lambda^2$ exceeds the level spacing, the normalized variance $\sigma_G^2/\sigCOE^2$ becomes essentially a function of $\lambda$ alone, independent of $D$ up to weak polynomial-in-$n$ corrections. It reaches the random-matrix value only when $\Gamma$ becomes comparable to the window width, at $\lambda$ of the order of the exchange coupling, while the level statistics of the same spectra are already Wigner--Dyson at $\lambda\propto D^{-1/2}$.

(iv)~The \emph{random-matrix endpoint}, derived exactly at finite $D$ (Theorem~\ref{thm:rmt}): $\sigCOE^2=f(1-f)D/[(D-1)(D+2)]$, which fixes the prefactor of the $e^{-S}$ estimate, including the completeness factor $(1-f)$.

The paper is organized as follows. Section~\ref{sec:model} defines the model, the channels, the two window prescriptions, and the numerical method. Section~\ref{sec:exact} derives the exact results. Section~\ref{sec:weak} develops the edge-resonance theory and compares it with exact diagonalization. Section~\ref{sec:crossover} analyzes the crossover to saturation and contrasts it with level statistics. Section~\ref{sec:discussion} discusses implications and limitations, and Sec.~\ref{sec:conclusion} concludes. All derivations are given in full in Appendices~\ref{app:edge}--\ref{app:golden}; numerical details and a gate-level implementation are in Appendices~\ref{app:numerics} and~\ref{app:circuit}.

% ======================================================================
\section{Model, channels, and windows}
\label{sec:model}

\subsection{Hamiltonian and charge}
We consider an open chain of $n$ spins-$\tfrac12$,
\begin{align}
H(\lambda)&=H_0+\lambda V,\qquad V=\sum_{i=1}^{n}S^x_i ,\nonumber\\
H_0&=\sum_{i=1}^{n-1}\bm S_i\!\cdot\!\bm S_{i+1}+\sum_{i=1}^{n}h_iS^z_i ,
\label{eq:H}
\end{align}
with $\bm S_i=\bm\sigma_i/2$ and independent random fields $h_i$ uniformly distributed in $[-W,W]$. $H_0$ is the random-field Heisenberg chain, the standard model of the ergodic-to-localized crossover~\cite{Pal:2010,Luitz:2015,Suntajs:2020}. We fix $W=0.5$, deep in the ergodic regime: the mean ratio of consecutive level spacings~\cite{Oganesyan:2007,Atas:2013} in the half-filling sector is $\langle r\rangle=0.530(2)$ at $n=12$ (Gaussian orthogonal ensemble, GOE: $0.5307$; Poisson: $0.3863$). The calibration curve $\langle r\rangle(W)$ is given in Appendix~\ref{app:numerics}.

The charge is the number of flipped spins,
\begin{equation}
\hat Q=\sum_{i=1}^{n}\Big(\tfrac12-S^z_i\Big),\qquad [H_0,\hat Q]=0 ,
\label{eq:Q}
\end{equation}
a genuine global $U(1)$ symmetry. Its sectors $\mathcal H_N$ have dimensions $D_N=\binom{n}{N}$. The transverse field $V$ changes $\hat Q$ by $\pm1$ and breaks the symmetry for $\lambda\neq0$; we use repeatedly the commutator
\begin{equation}
\hat C\equiv[V,\hat Q]=i\sum_{i=1}^{n}S^y_i ,
\label{eq:C}
\end{equation}
which follows from $[S^x,S^z]=-iS^y$. $H(\lambda)$ is real symmetric in the computational basis (time-reversal invariant), so the relevant random-matrix class is orthogonal. The full Hilbert space has dimension $D=2^n$ and we define $S\equiv\ln D$.

\subsection{Channels and observable}
The channels are real Haar-random unit vectors supported in fixed charge sectors,
\begin{equation}
|A\rangle\in\mathcal H_{N_A},\qquad |B\rangle\in\mathcal H_{N_B},\qquad N_B=N_A+\Delta Q ,
\label{eq:channels}
\end{equation}
with $N_A=n/2$ and $\Delta Q\in\{1,2\}$. Generic channels avoid any accidental structure of particular product states; we comment on fixed channels in Appendix~\ref{app:numerics}. Let $H(\lambda)|I\rangle=E_I|I\rangle$ with $E_1<E_2<\dots<E_D$, and let $\Pw=\sum_{I\in\mathcal W}|I\rangle\langle I|$. The observable is Eq.~\eqref{eq:G}; Fig.~\ref{fig:setup}(a) shows the charge-resolved spectrum and a window. We consider two prescriptions for $\mathcal W$, with the same mean fraction $f=\tfrac12$:
\begin{itemize}
\item \emph{count window}: the central $r=fD$ eigenstates of $H(\lambda)$, i.e., a fixed number of microstates, as presupposed by the random-walk estimate;
\item \emph{energy window}: all eigenstates with $E_-\le E_I\le E_+$, where $E_\pm$ are fixed numbers (the central-$f$ quantiles of the pooled $\lambda=0$ spectrum over many disorder realizations), independent of $\lambda$ and of the realization, i.e., a microcanonical energy shell.
\end{itemize}

\subsection{Ensembles and estimators}
We diagonalize $H(\lambda)$ exactly in the full $2^n$-dimensional space for $n=8$, $10$, and $12$ ($D=256$, $1024$, $4096$), for up to $16$ values of $\lambda\in[0,2]$. For every disorder realization we compute the exact average of $G^2$ over Haar channels,
\begin{equation}
\mathcal T\equiv\Expec_{A,B}\big[G^2\big]=\frac{\Tr\big(\hat\Pi_A\Pw\hat\Pi_B\Pw\big)}{D_{N_A}D_{N_B}} ,
\label{eq:T}
\end{equation}
where $\hat\Pi_N$ projects onto $\mathcal H_N$ (derivation in Appendix~\ref{app:channel}), and we also sample $G$ for fresh Haar channel pairs to obtain typical values. The ensemble variance is $\sigma_G^2\equiv\Expec_h[\mathcal T]$. Ensemble sizes are $4000$ ($n=8$) and $1500$ ($n=10$) realizations per $\lambda$ in the weak-breaking regime and $30$ for $n=12$ ($\lambda\ge10^{-2}$), with smaller ensembles where fluctuations are small (Appendix~\ref{app:numerics}). Using Eq.~\eqref{eq:T} instead of a few fixed channel pairs removes channel noise entirely; the remaining fluctuations are disorder fluctuations, which become heavy tailed at small $\lambda$ for the reason explained in Sec.~\ref{sec:weak}.

% ======================================================================
\section{Exact structure}
\label{sec:exact}

\subsection{Superselection and the selection rule}
At $\lambda=0$ every eigenstate of $H_0$ can be chosen with definite charge, so $\Pw$ is block diagonal and $G=0$ identically in every realization. Numerically, $\max|G|<5\times10^{-16}$ over all realizations at $\lambda=0$.

For $\lambda\neq0$, expanding the Riesz projector $\Pw=(2\pi i)^{-1}\oint_\Gamma dz\,(z-H)^{-1}$ in powers of $\lambda$ gives (Appendix~\ref{app:dk})
\begin{align}
G&=\sum_{k\ge\Delta Q}\lambda^k G^{(k)},\nonumber\\
G^{(k)}&=\!\!\sum_{\alpha,\gamma_1,\dots,\gamma_{k-1},\beta}\!\! a_\alpha b_\beta\,V_{\alpha\gamma_1}V_{\gamma_1\gamma_2}\cdots V_{\gamma_{k-1}\beta}\nonumber\\
&\qquad\qquad\times\chi_{\mathcal W}[e_\alpha,e_{\gamma_1},\dots,e_{\gamma_{k-1}},e_\beta],
\label{eq:DK}
\end{align}
where $e_\alpha$ and $|\alpha\rangle$ are the eigenvalues and charge-resolved eigenvectors of $H_0$, $a_\alpha=\langle A|\alpha\rangle$, $b_\beta=\langle\beta|B\rangle$, and $\chi_{\mathcal W}[x_0,\dots,x_k]$ is the $k$th divided difference of the indicator function of $\mathcal W$ (the Daleckii--Krein structure~\cite{Daleckii:1965,Bhatia:1997}). Because $V$ changes the charge by one unit, a chain from $\mathcal H_{N_A}$ to $\mathcal H_{N_B}$ needs at least $\Delta Q$ steps, which is the selection rule $|G|_{\rm typ}\propto\lambda^{\Delta Q}$. Moreover, a divided difference of order $k\geq1$ of a function that is constant on all its nodes vanishes, so every term in Eq.~\eqref{eq:DK} requires the chain of energies to straddle a window edge. The next result makes this edge localization exact.

\subsection{Edge representation}

\begin{theorem}[Edge representation]
\label{thm:edge}
Let $[H_0,\hat Q]=0$, $H=H_0+\lambda V$ with nondegenerate spectrum, and let $|A\rangle$, $|B\rangle$ be charge eigenstates with $N_A\neq N_B$. Then, for any set $\mathcal W$ of eigenstates,
\begin{align}
G=\frac{1}{N_A-N_B}&\sum_{I\in\mathcal W}\sum_{J\notin\mathcal W}
\Big[\langle A|J\rangle\,Q_{JI}\,\langle I|B\rangle\nonumber\\
&\qquad\qquad-\langle A|I\rangle\,Q_{IJ}\,\langle J|B\rangle\Big],\nonumber\\
Q_{IJ}=\frac{\lambda\,C_{IJ}}{E_I-E_J},&\qquad C_{IJ}=\langle I|\hat C|J\rangle .
\label{eq:edge}
\end{align}
\end{theorem}

The proof (Appendix~\ref{app:edge}) uses only $\langle A|[\hat Q,\Pw]|B\rangle=(N_A-N_B)G$ and the exact relation $(E_I-E_J)Q_{IJ}=\lambda C_{IJ}$ obtained from $[H,\hat Q]=\lambda\hat C$. We have verified Eq.~\eqref{eq:edge} numerically to a relative accuracy of $10^{-10}$ for $n\le10$ and $\lambda$ from $10^{-3}$ to $1$. Three consequences organize the rest of the paper. First, symmetry violation is generated only by pairs $(I,J)$ with one member inside and one outside the window, weighted by $1/(E_I-E_J)$; pairs deep inside the window cancel by completeness. Second, the weights are largest for nearly degenerate pairs at the edges, which is the origin of the resonances analyzed in Sec.~\ref{sec:weak}. Third, if the sharp projector is replaced by a smooth function $F(H)$, the factor $(p_J-p_I)/(E_I-E_J)$ [$p_I=1$ for $I\in\mathcal W$] becomes $[F(E_J)-F(E_I)]/(E_I-E_J)$, which is bounded by $\max|F'|$: the small denominators, and with them the resonances, are a property of sharp window edges.

\subsection{Exact channel average}
For independent real Haar channels in their sectors, $\Expec_{A,B}[G]=0$ and Eq.~\eqref{eq:T} holds exactly (Appendix~\ref{app:channel}). Using completeness, $\sum_J(\hat\Pi_A)_{IJ}(\hat\Pi_B)_{JI}=(\hat\Pi_A\hat\Pi_B)_{II}=0$, the trace can be written in edge form as well,
\begin{equation}
\mathcal T=-\frac{1}{D_{N_A}D_{N_B}}\sum_{I\in\mathcal W}\sum_{J\notin\mathcal W}(\hat\Pi_A)_{IJ}(\hat\Pi_B)_{JI} ,
\label{eq:Tedge}
\end{equation}
which we verified to $10^{-14}$. Equations~\eqref{eq:edge} and~\eqref{eq:Tedge} are exact for every realization and every $\lambda$.

\subsection{The random-matrix endpoint}
If the window eigenvectors were Haar random in the full space, Eq.~\eqref{eq:T} would be fixed by the second moment of a random projector.

\begin{theorem}[Random-projector variance]
\label{thm:rmt}
Let $\hat P$ project onto a Haar-random $r$-dimensional subspace of $\mathbb R^D$ (orthogonal class) or $\mathbb C^D$ (unitary class), $f=r/D$, and let $|a\rangle\perp|b\rangle$ be fixed unit vectors. Then $\Expec\langle a|\hat P|b\rangle=0$ and
\begin{equation}
\Expec\big|\langle a|\hat P|b\rangle\big|^2=
\begin{cases}
\sigCOE^2\equiv \dfrac{f(1-f)\,D}{(D-1)(D+2)} & \text{(orthogonal)},\\[3mm]
\dfrac{f(1-f)\,D}{D^2-1} & \text{(unitary)}.
\end{cases}
\label{eq:rmt}
\end{equation}
\end{theorem}

Two independent proofs, one from rotational invariance alone and one from the Beta distribution of $\langle a|\hat P|a\rangle$~\cite{Zyczkowski:2001}, are given in Appendix~\ref{app:rmt}; Monte Carlo sampling confirms Eq.~\eqref{eq:rmt} to within $0.5\%$ statistical accuracy (Table~\ref{tab:mc}). Inserting Eq.~\eqref{eq:rmt} into Eq.~\eqref{eq:T}, every one of the $D_{N_A}D_{N_B}$ cross elements contributes $\sigCOE^2$, so the channel-averaged variance equals $\sigCOE^2$ independently of the sectors. With $D=e^{S}$,
\begin{equation}
\sigCOE=\sqrt{f(1-f)}\;e^{-S/2}\,\big[1+O(e^{-S})\big].
\label{eq:rmtasym}
\end{equation}
This is the random-walk estimate with its prefactor fixed. The factor $(1-f)$ expresses completeness: since $\langle A|B\rangle=0$, the contributions of all $D$ eigenstates sum to zero, and $G\to0$ as the window fills the spectrum. The question addressed next is how, and at what breaking strength, the physical $H(\lambda)$ approaches this value.

% ======================================================================
\section{Weak breaking: edge resonances}
\label{sec:weak}

\begin{figure*}[t]
\centering
\includegraphics[width=\textwidth]{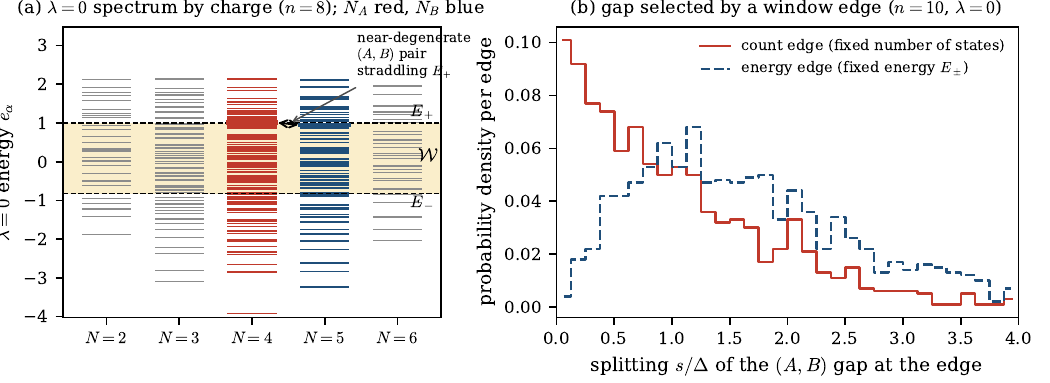}
\caption{\label{fig:setup}Setup and the gap-selection mechanism. (a)~Spectrum of $H_0$ for one disorder realization ($n=8$), resolved by charge sector. The count window $\mathcal W$ (shaded) holds the central half of all $D=256$ levels; levels of the channel sectors $N_A=4$ (red) and $N_B=5$ (blue) interleave without repulsion. The arrow marks the closest $(A,B)$ pair straddling the upper edge. (b)~Distribution of the splitting $s$ of the gap selected by a window edge, restricted to gaps bounded by one level of each channel sector ($n=10$, $4000$ realizations, $s$ in units of the local mean spacing $\Delta$). A count edge (between the $r$th and $(r+1)$th level) selects gaps without bias and the density stays finite as $s\to0$; an energy edge (a fixed energy) selects a gap with probability proportional to its size and the density vanishes linearly.}
\end{figure*}

\subsection{Typical amplitude}
\begin{figure*}[t]
\centering
\includegraphics[width=\textwidth]{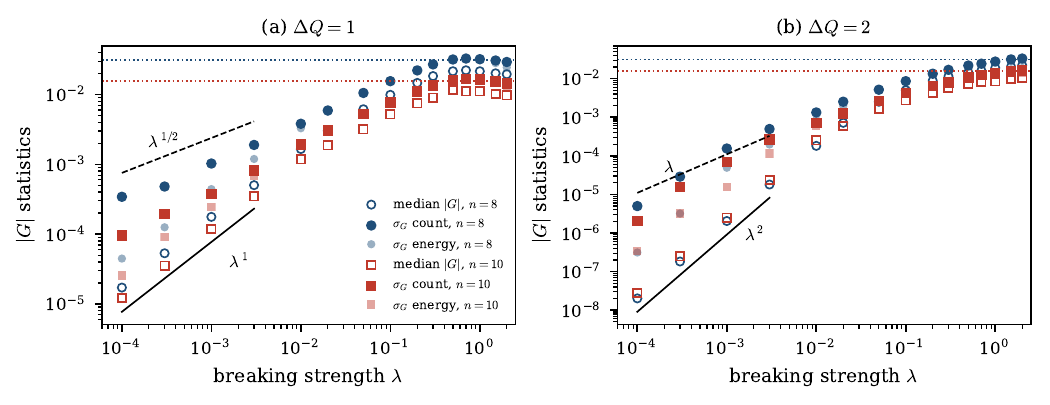}
\caption{\label{fig:selection}Typical versus root-mean-square violation. Open symbols: median of $|G|$; filled symbols: $\sigma_G$ for the count window (dark) and the energy window (light), for $n=8$ (circles) and $n=10$ (squares). Solid lines: selection rule $\lambda^{\Delta Q}$. Dashed lines: resonant law $\sigma_G\propto\lambda^{\Delta Q/2}$ of the count window. Dotted horizontal lines: the random-matrix value $\sigma_{\rm RMT}$ of Eq.~\eqref{eq:rmt}. At $\lambda=0$, $|G|<5\times10^{-16}$ (not shown).}
\end{figure*}

To lowest order, Eq.~\eqref{eq:DK} gives for $\Delta Q=1$
\begin{equation}
G^{(1)}=\sum_{\alpha\in A}\sum_{\beta\in B}a_\alpha b_\beta V_{\alpha\beta}\,\frac{\chi_{\mathcal W}(e_\alpha)-\chi_{\mathcal W}(e_\beta)}{e_\alpha-e_\beta} ,
\label{eq:G1}
\end{equation}
a sum over $(A,B)$ pairs on opposite sides of an edge. It is a sum of many terms with random signs, so its typical size is set by the bulk of non-resonant pairs and scales as $\lambda$; for $\Delta Q=2$ the leading term is $\lambda^2G^{(2)}$. The median of $|G|$ over disorder and Haar channels [Fig.~\ref{fig:selection}] follows the selection rule with fitted exponents $0.99\pm0.01$ for $\Delta Q=1$ and $1.99\pm0.02$ for $\Delta Q=2$ over $\lambda\in[10^{-4},3\times10^{-3}]$, for both window prescriptions and for $n=8$ and $10$. The first-order formula~\eqref{eq:G1} agrees with exact diagonalization to relative accuracy $1.1\times10^{-8}$ at $\lambda=10^{-6}$ (Appendix~\ref{app:numerics}).

\subsection{Two-level resonances at the edge}
The variance does not follow the typical value. Equation~\eqref{eq:G1} contains denominators $e_\alpha-e_\beta$ that are not protected: at $\lambda=0$ the sectors are independent, levels of different charge do not repel, and the density of $(A,B)$ pairs with splitting $s$ is finite as $s\to0$ [Fig.~\ref{fig:setup}(a)]. For a pair with $|s|\lesssim\lambda|V_{\alpha\beta}|$ perturbation theory fails and the pair must be treated exactly. Diagonalizing the $2\times2$ block $\big(\begin{smallmatrix} e_\alpha & g\\ g & e_\beta\end{smallmatrix}\big)$, $g=\lambda V_{\alpha\beta}$, gives eigenvectors $\cos\theta|\alpha\rangle\pm\sin\theta|\beta\rangle$ (up to ordering) with $\tan2\theta=2g/s$. If one of them lies inside $\mathcal W$ and the other outside, the pair contributes to $\langle\alpha|\Pw|\beta\rangle$ the amount
\begin{equation}
x=\cos\theta\sin\theta=\frac{g}{\sqrt{s^2+4g^2}} ,
\label{eq:x}
\end{equation}
which reduces to the first-order term $g/s$ for $|s|\gg|g|$ and saturates at $|x|=\tfrac12$ on resonance (Appendix~\ref{app:resonance}). If both lie inside, their contributions cancel exactly, as required by Theorem~\ref{thm:edge}. Averaging over Haar channels, each straddling pair contributes $x^2/(D_{N_A}D_{N_B})$ to $\sigma_G^2$.

The ensemble average of $x^2$ is controlled by the probability that a nearly degenerate $(A,B)$ pair sits at an edge, and this is where the two window prescriptions differ (Appendix~\ref{app:resonance}):

\emph{Count window.} The edge lies in the gap between the $r$th and $(r+1)$th levels, so it selects a \emph{gap}, uniformly among gaps. The probability that this gap is an $(A,B)$ gap of splitting $|s|<\epsilon$ is $\nu\epsilon$ for small $\epsilon$, with
\begin{equation}
\nu=\sum_{e=\pm}\frac{2\rho_A(E_e)\rho_B(E_e)}{\rho(E_e)} ,
\label{eq:nu}
\end{equation}
where $\rho_{A,B}$ are the densities of states of the channel sectors, $\rho$ the total density, and the sum runs over the two edges. Integrating Eq.~\eqref{eq:x} against this finite density,
\begin{equation}
\sigma_G^2\simeq\frac{\nu}{D_{N_A}D_{N_B}}\int_0^\infty\!ds\,\overline{\frac{g^2}{s^2+4g^2}}
=\frac{\pi}{4}\,\frac{\nu\,\overline{|V|}}{D_{N_A}D_{N_B}}\,\lambda ,
\label{eq:count}
\end{equation}
where the bar averages over the distribution of near-degenerate cross-sector matrix elements. The variance is linear in $\lambda$ and dominated by rare resonant realizations.

\emph{Energy window.} A fixed energy $E_e$ falls inside a gap with probability proportional to the gap size---the inspection paradox of renewal theory~\cite{Feller:1971}---so the density of selected $(A,B)$ gaps vanishes as $s\to0$ [Fig.~\ref{fig:setup}(b)]. A pair with midpoint $m$ contributes when exactly one hybridized level lies in $\mathcal W$, i.e., when $|m-E_e|<\tfrac12\sqrt{s^2+4g^2}$, so
\begin{align}
\sigma_G^2&\simeq\sum_{e=\pm}\frac{\rho_A\rho_B}{D_{N_A}D_{N_B}}\int\! ds\,\overline{\frac{g^2}{\sqrt{s^2+4g^2}}}\nonumber\\
&\simeq\sum_{e=\pm}\frac{2\rho_A\rho_B\overline{V^2}}{D_{N_A}D_{N_B}}\,\lambda^2\ln\frac{\omega_c}{\lambda},
\label{eq:energy}
\end{align}
with $\omega_c$ the frequency scale beyond which the off-diagonal matrix elements of $V$ decay. Resonances now contribute only a logarithm.

For $\Delta Q=1$ the two prescriptions therefore give
\begin{equation}
\sigma_G^2\propto
\begin{cases}
\lambda & \text{(count)},\\
\lambda^2\ln(1/\lambda) & \text{(energy)},
\end{cases}
\qquad |G|_{\rm typ}\propto\lambda ,
\label{eq:laws}
\end{equation}
so that the root-mean-square and typical violations part ways parametrically, $\sigma_G/|G|_{\rm typ}\propto\lambda^{-1/2}$, only for the count window. For $\Delta Q=2$ the resonant coupling is the second-order effective matrix element $g_{\rm eff}\propto\lambda^2$ through the intermediate sector, and the same argument gives $\sigma_G^2\propto\lambda^2$ (count) and $\lambda^4$ (energy), up to logarithms. Because a resonance now occurs with probability $\propto\lambda^2$ per realization, direct ensembles undersample it at small $\lambda$; the fitted variance exponents at $n=8$ over $10^{-4}\le\lambda\le3\times10^{-3}$ are $2.7$ and $3.9$ for the count and energy windows. The resonant law is nevertheless confirmed directly through its tail, which follows Eq.~\eqref{eq:tail} with $\nu\to\nu_2$ and $\overline{|V|}\lambda\to\overline{|g_{\rm eff}|}$, both computed from $\lambda=0$ data (Appendix~\ref{app:resonance}, Fig.~\ref{fig:dq2}): at $y=0.1$ the measured $P/\lambda^2$ is $150\pm31$, $93\pm10$, $67\pm4$ for $\lambda=0.002,\,0.005,\,0.01$, against the predictions $152$, $105$, $76$.

\subsection{Comparison with exact diagonalization}
\begin{figure*}[t]
\centering
\includegraphics[width=\textwidth]{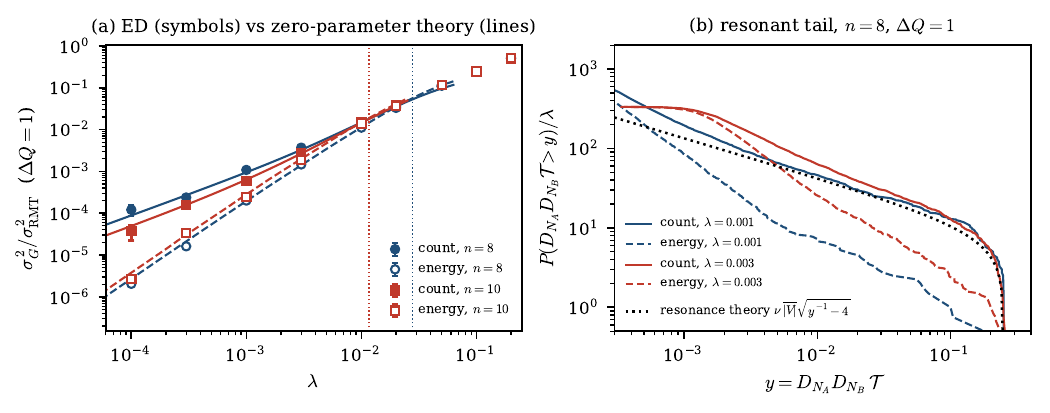}
\caption{\label{fig:resonance}Edge-resonance theory versus exact diagonalization ($\Delta Q=1$). (a)~Normalized variance $\sigma_G^2/\sigma_{\rm RMT}^2$ for the count window (filled) and energy window (open), $n=8$ (circles) and $n=10$ (squares), with bootstrap errors. Lines: resonance theory evaluated from $\lambda=0$ data only, solid for count and dashed for energy windows, plotted up to $2.5\,\lambda_\Delta$; vertical dotted lines mark $\lambda_\Delta$ [Eq.~\eqref{eq:lD}]. (b)~Complementary distribution of the channel-averaged variance of single realizations, rescaled by $\lambda$, for $n=8$. Count-window data at two values of $\lambda$ collapse onto the parameter-free prediction~\eqref{eq:tail} (dotted); energy-window tails (dashed) are parametrically smaller.}
\end{figure*}

Equations~\eqref{eq:count} and~\eqref{eq:energy} are the leading terms of a resonance theory that can be evaluated completely from $\lambda=0$ data: the charge-resolved spectra, the matrix elements $V_{\alpha\beta}$ between eigenstates of neighboring sectors, and the densities of states. We sum $x^2$ over all straddling $(A,B)$ pairs with $|s|$ larger than a quarter of the local level spacing directly from the sampled spectra, and integrate the rare near-degenerate pairs analytically with the exact $s\to0$ densities of Eqs.~\eqref{eq:count} and~\eqref{eq:energy} (Appendix~\ref{app:resonance}). No quantity is fitted to the $\lambda>0$ data. Figure~\ref{fig:resonance}(a) compares the result with exact diagonalization of the full $2^n$-dimensional problem. For $n=8$ the theory reproduces the count-window variance within $18\%$ and the energy-window variance within $6\%$ over $10^{-3}\le\lambda\le5\times10^{-2}$; the same holds for $n=10$ (within $12\%$ and $13\%$, respectively). At the two smallest values of $\lambda$ the energy-window data fall $\approx24\%$ below the theory. This is an estimator effect, not a failure of the theory: the logarithm in Eq.~\eqref{eq:energy} receives equal weight from every scale of $s$ down to $\lambda|V|$, while a finite ensemble only samples splittings $s\gtrsim(\rho_A\rho_B N_{\rm real})^{-1/2}$. On identical realizations, by contrast, the pair sum reproduces the exact $\mathcal T$ with a median relative deviation below $10^{-4}$ at $\lambda=10^{-4}$ (Appendix~\ref{app:numerics}).

The resonant tail itself is a sharp, parameter-free test. Within the two-level reduction, the channel-averaged variance of a realization with a resonant edge pair is $\mathcal T=x^2/(D_{N_A}D_{N_B})$, and Eq.~\eqref{eq:x} with a flat density of splittings gives
\begin{equation}
P\big(D_{N_A}D_{N_B}\mathcal T>y\big)\simeq\nu\,\overline{|V|}\,\lambda\,\sqrt{y^{-1}-4},\qquad 0<y<\tfrac14 ,
\label{eq:tail}
\end{equation}
for $y$ above the non-resonant background. Figure~\ref{fig:resonance}(b) shows that the count-window distributions at $\lambda=10^{-3}$ and $3\times10^{-3}$, rescaled by $\lambda$, collapse onto Eq.~\eqref{eq:tail} including the cutoff at $y=\tfrac14$, while the energy-window tails are suppressed as predicted.

The resonant coefficient of the count window, $\sigma_G^2/\sigCOE^2=\tilde c_1\lambda$ with $\tilde c_1=(\pi/4)\nu\overline{|V|}/(D_{N_A}D_{N_B}\sigCOE^2)$, decreases with system size: $\tilde c_1=0.88$, $0.48$, and $0.27$ for $n=8$, $10$, and $12$. The scaling $\tilde c_1\propto D^{-1/2}$ follows from eigenstate thermalization: the cross-sector matrix elements of the local operator $V$ are of order $\overline{|V|}\propto\rho_B^{-1/2}$, while $\nu/(D_{N_A}D_{N_B})\propto\sigCOE^2/\sigma_E$ with $\sigma_E$ the spectral width. The resonant regime therefore shrinks with $D$, which leads to the crossover analyzed next.

% ======================================================================
\section{Crossover to random-matrix saturation}
\label{sec:crossover}

\subsection{Golden-rule broadening and two breaking scales}
The two-level reduction applies while resonances are isolated. Each $\lambda=0$ eigenstate acquires a golden-rule decay width into the neighboring sectors~\cite{Harney:1986,Mallayya:2019},
\begin{equation}
\Gamma_\alpha=2\pi\lambda^2\!\!\sum_{\beta:\,N_\beta=N_\alpha\pm1}\!\!|V_{\alpha\beta}|^2\,\delta(e_\alpha-e_\beta)\equiv\gamma_\alpha\lambda^2 ,
\label{eq:Gamma}
\end{equation}
the spreading width of nuclear physics. Comparing $\Gamma$ with the two energy scales of the problem defines two breaking strengths,
\begin{equation}
\lD=\sqrt{\Delta/\gamma_0},\qquad \lB=\sqrt{\delta E_{\mathcal W}/\gamma_0} ,
\label{eq:lD}
\end{equation}
where $\gamma_0$ is the mean of $\gamma_\alpha$ at the window edges, $\Delta$ the mean level spacing there, and $\delta E_{\mathcal W}=E_+-E_-$ the window width. All three are computed from $\lambda=0$ data (Table~\ref{tab:params}). Since $\Delta\propto1/D$ while $\gamma_0$ and $\delta E_{\mathcal W}$ are intensive up to factors polynomial in $n$, $\lD\propto D^{-1/2}$ decreases exponentially with system size whereas $\lB$ does not.

\begin{table}[b]
\caption{\label{tab:params}Parameters of the resonance theory, all from $\lambda=0$ data ($W=0.5$, $f=\tfrac12$, count window). $\Delta$: mean level spacing at the edges; $\gamma_0$: golden-rule coefficient, Eq.~\eqref{eq:Gamma}; $\delta E_{\mathcal W}$: window width; $\tilde c_1$: resonant coefficient, $\sigma_G^2/\sigCOE^2=\tilde c_1\lambda$ for $\lambda\ll\lD$.}
\begin{ruledtabular}
\begin{tabular}{cccccccc}
$n$ & $D$ & $\Delta$ & $\gamma_0$ & $\delta E_{\mathcal W}$ & $\lD$ & $\lB$ & $\tilde c_1$\\
\colrule
8 & 256 & $1.61\times10^{-2}$ & 20.8 & 1.77 & 0.0278 & 0.292 & 0.878\\
10 & 1024 & $4.62\times10^{-3}$ & 33.8 & 1.97 & 0.0117 & 0.241 & 0.479\\
12 & 4096 & $1.20\times10^{-3}$ & 45.2 & 2.13 & 0.0052 & 0.217 & 0.270\\
\end{tabular}
\end{ruledtabular}
\end{table}

For $\lambda\gtrsim\lD$ the eigenstates of $H(\lambda)$ are Breit--Wigner superpositions of $\sim\Gamma/\Delta$ unperturbed states of neighboring charge~\cite{Wigner:1955,BohrMottelson:1969,Flambaum:1997}; resonances overlap and the distinction between the two window prescriptions disappears [Fig.~\ref{fig:resonance}(a)]. The energy denominators are cut off by $\Gamma$ instead of by $g$, $x^2\to g^2/(s^2+4g^2+\Gamma^2)$, and the continuum limit of the edge sum gives
\begin{equation}
\frac{\sigma_G^2}{\sigCOE^2}\simeq\frac{D}{f(1-f)\,D_{N_B}}\sum_{e=\pm}\varrho_A(E_e)\,\frac{\Gamma_{A\to B}}{\pi}\,\ln\frac{\omega_c}{\Gamma}
\label{eq:golden}
\end{equation}
for $\lD\lesssim\lambda\ll\lB$, where $\varrho_A=\rho_A/D_{N_A}$ is the normalized density of states of sector $N_A$ and $\Gamma_{A\to B}\propto\lambda^2$ the partial width into sector $N_B$ (Appendix~\ref{app:golden}). No factor on the right-hand side depends exponentially on $n$: $\varrho_A$ is of order the inverse spectral width, $\Gamma_{A\to B}\propto n\lambda^2$ because $V$ is a sum of $n$ local terms, and $D/D_{N_B}$, the inverse weight of one charge sector, grows as $\sqrt n$. Thus, once resonances overlap, the normalized variance is essentially a function of $\lambda$ alone, independent of the Hilbert-space dimension, and it approaches unity only when $\Gamma$ reaches the window width, at $\lambda\sim\lB$. With $\Gamma$ included, the resonance theory tracks exact diagonalization up to $\lambda\approx2\lD$ [Fig.~\ref{fig:resonance}(a)]; beyond that, the processes it omits---broadening into further sectors and the redistribution of spectral weight across the whole window---take over.

\subsection{Size independence and the contrast with level statistics}
\begin{figure*}[t]
\centering
\includegraphics[width=\textwidth]{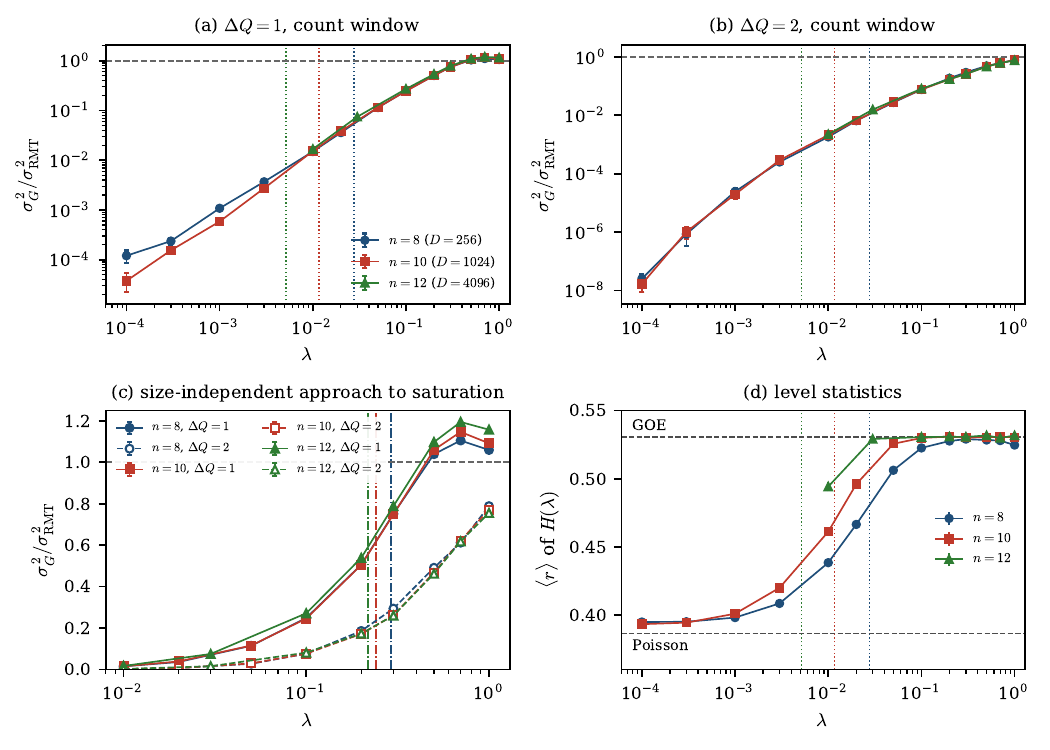}
\caption{\label{fig:crossover}Crossover from superselection to random-matrix saturation (count window). (a,b)~Normalized variance $\sigma_G^2/\sigma_{\rm RMT}^2$ for $\Delta Q=1$ and $2$ and $n=8$, $10$, $12$; vertical dotted lines mark $\lambda_\Delta$ for each size [Eq.~\eqref{eq:lD}]. (c)~The same on a linear scale for $\lambda\ge10^{-2}$ (filled: $\Delta Q=1$, open: $\Delta Q=2$); dash-dotted lines mark $\lambda_{\mathcal W}$. (d)~Mean gap ratio $\langle r\rangle$ of the full spectrum of $H(\lambda)$ (central $80\%$ of levels). Level statistics become Wigner--Dyson near $\lambda_\Delta\propto D^{-1/2}$, whereas the symmetry-violating variance is size independent beyond $\lambda_\Delta$ and saturates only for $\lambda\gtrsim\lambda_{\mathcal W}$.}
\end{figure*}

Figure~\ref{fig:crossover} shows the full crossover for $n=8$, $10$, and $12$. Three regimes are visible. For $\lambda\lesssim\lD$ the curves are ordered by size, with the resonant coefficient $\tilde c_1\propto D^{-1/2}$. For $\lambda\gtrsim\lD$ they merge: at $\lambda\ge0.03$ the normalized variances of $n=8$, $10$, and $12$ agree within $10\%$, although $D$ changes by a factor of $16$; the small residual drift is of the polynomial-in-$n$ type anticipated by Eq.~\eqref{eq:golden}, whereas in the resonant regime the curves differ by factors $\propto D^{-1/2}$. The approach to the random-matrix value is slow and size independent [Fig.~\ref{fig:crossover}(c)]; for $\Delta Q=1$ the count-window variance reaches $\sigma_G^2/\sigCOE^2=1.11\text{--}1.19$ at $\lambda=0.7$, and for $\Delta Q=2$, which requires two golden-rule steps, it reaches only $0.61\text{--}0.62$ at the same $\lambda$.

The level statistics of the same spectra behave very differently [Fig.~\ref{fig:crossover}(d)]. At $\lambda=0$ the full spectrum is a superposition of $n+1$ independent sectors and $\langle r\rangle\approx0.39$, close to the Poisson value. The gap ratio rises to the GOE value around $\lD$, i.e., at breaking strengths that shrink with system size, in agreement with the criterion that the rms coupling be comparable to the level spacing~\cite{Guhr:1990,Weidenmuller:2009} and with the size scaling of the onset of chaos in perturbed many-body systems~\cite{Rabson:2004,LeBlond:2021}. At $\lambda=0.1$, for example, $\langle r\rangle=0.523, 0.530, 0.531$ for $n=8$, $10$, $12$, while $\sigma_G^2/\sigCOE^2=0.25, 0.25, 0.27$. Wigner--Dyson level statistics therefore do not imply that a weakly broken symmetry is maximally violated in the eigenstates. Between $\lD$ and $\lB$ the eigenstates are extended across charge sectors only on the energy scale $\Gamma$, the analog of the non-ergodic extended phase of generalized Rosenzweig--Porter ensembles~\cite{Kravtsov:2015,DeTomasi:2019}, in which local level statistics are already Wigner--Dyson while eigenvectors are not yet ergodic.

\subsection{The saturated regime}
For $\lambda\gtrsim\lB$ the count-window variance for $\Delta Q=1$ is within $19\%$ of the random-matrix value for all three sizes. The residual deviations, of either sign, are non-universal: they reflect the energy dependence of the charge-relaxation spectral function relative to the window, which the theorem assumes to be flat, and at $\lambda\gtrsim1.5$ the transverse field begins to dominate $H$ and $\langle r\rangle$ decreases again (for $n=8$, $\langle r\rangle=0.473$ at $\lambda=2$). We therefore regard Eq.~\eqref{eq:rmt} as the endpoint of a crossover that a physical Hamiltonian reaches over a finite range of $\lambda$, not as a property of any small breaking. Two practical remarks follow (Appendix~\ref{app:numerics}). First, for fixed rather than Haar channels, $\sigma_G$ depends strongly on the channel even at saturation ($\sigma_G/\sigCOE$ between 0.49 and 1.21 for six fixed pairs at $n=8$, $\lambda=0.7$), because the disorder average does not randomize a channel's overlap with the energy shell; estimates from a few fixed channel pairs are therefore unreliable, and the exact channel average~\eqref{eq:T} should be used. Second, at small $\lambda$ the count-window estimator is limited by the resonance rate: a faithful sample of $\sigma_G^2$ needs $N_{\rm real}\gg1/(\nu\overline{|V|}\lambda)$ realizations for $\Delta Q=1$ and $N_{\rm real}\gg1/(\nu\overline{|g_{\rm eff}|})$ with $g_{\rm eff}\propto\lambda^2$ for $\Delta Q=2$.

% ======================================================================
\section{Discussion}
\label{sec:discussion}

\emph{What the random-walk estimate captures.} In a chaotic system with a weakly broken $U(1)$ symmetry, the charge-violating microcanonical amplitude does reach the random-matrix magnitude $e^{-S/2}$, with the exact finite-$D$ prefactor $\sqrt{f(1-f)D^2/[(D-1)(D+2)]}$ of Theorem~\ref{thm:rmt}. The factor $(1-f)$ is a completeness effect absent from the naive random walk; it is the reason the amplitude vanishes when the window covers the whole spectrum. This is the sense in which the estimate is correct: it describes the endpoint.

\emph{What it misses.} The endpoint is reached only when the breaking-induced spreading width becomes comparable to the energy window, $\lambda\sim\lB$. Before that, the amplitude is generated exclusively at the window edges (Theorem~\ref{thm:edge}), its rms value is suppressed relative to the random-matrix value by a factor that depends on $\lambda$ but not on $D$ [Eq.~\eqref{eq:golden}], and in the regime of isolated resonances its statistics are heavy tailed, with an rms value that depends on whether the window counts states or fixes energies [Eq.~\eqref{eq:laws}]. None of this structure is visible in an estimate that presupposes random eigenvectors. In particular, the concern that the random walk might be invalidated by ``extra cancellations''~\cite{Zhiboedov:notes} is answered in a precise way: there are cancellations---all pairs inside the window cancel by completeness---but their effect is to localize the violation at the edges, not to suppress it below the random-matrix value at saturation.

\emph{Relation to black-hole physics.} Our model contains no gravity, and $S=\ln D$ is a Hilbert-space dimension, not a horizon area. The results concern any Hamiltonian with a weakly broken $U(1)$ symmetry whose symmetric part is chaotic within each sector. If black-hole microstates are described by such statistics~\cite{Cotler:2017,Saad:2019,Belin:2022}, our analysis suggests that an estimate $|\mathcal A|^2\sim e^{-S}$ for a charge-violating amplitude~\cite{Bah:2023} carries an implicit assumption about the strength of the violating dynamics relative to the energy resolution of the process, and about how the microcanonical ensemble is specified. We do not claim more than this.

\emph{Relation to symmetry breaking in nuclei and random-matrix models.} Spreading widths, the static and dynamical criteria for isospin breaking~\cite{Harney:1986}, and the crossover of spectral statistics~\cite{Guhr:1990,French:1988} are classical. What is new here is the observable: a matrix element of a spectral projector with sharp edges between channels of different charge. Its edge representation makes the role of the window explicit, explains why the crossover of this observable is controlled by $\Gamma/\delta E_{\mathcal W}$ rather than $\Gamma/\Delta$, and reveals the dependence on the window prescription.

\emph{Quantum simulation.} The charge of Eq.~\eqref{eq:Q} is diagonal in the computational basis, so the charge-sector populations of an evolved state are directly measurable, and at $\lambda=0$ any measured leakage between sectors is an error event~\cite{BonetMonroig:2018,McArdle:2019}. Appendix~\ref{app:circuit} gives the second-order Trotter circuit for $e^{-iH(\lambda)t}$, validates it against exact evolution with Qiskit Aer, and documents a simulation pitfall that hides all Trotter error. Accessing $G$ itself requires an energy filter such as quantum phase estimation~\cite{Kitaev:1995} or a polynomial spectral filter~\cite{Gilyen:2019}. Our results imply a clear hierarchy of costs: the size-independent golden-rule and saturated regimes require an energy resolution of order $\Gamma$, which is polynomial in $n$, whereas the resonant regime requires resolving individual levels near the edges, which takes a circuit depth of the order of the Heisenberg time $\propto D$.

\emph{Limitations.} Our sizes are limited to $D\le4096$ in the full Hilbert space. The resonance theory is evaluated up to $n=12$ from $\lambda=0$ data, and the size independence beyond $\lD$ is established for $n=10$ and $12$. The breaking operator is a uniform field; a random or local breaking changes $\gamma_0$ but not the structure of Theorem~\ref{thm:edge}. The window fraction is $f=\tfrac12$ throughout; the theory applies to any $f$, with Eq.~\eqref{eq:rmt} vanishing as $f\to1$.

% ======================================================================
\section{Conclusion}
\label{sec:conclusion}

We have analyzed the statistics of the charge-violating microcanonical amplitude $G=\langle A|\Pw|B\rangle$ in a chaotic spin chain with a weakly broken $U(1)$ symmetry, from exact conservation to complete mixing. An exact edge representation shows that the violation is generated only by eigenstate pairs straddling the window edges. At weak breaking the typical amplitude obeys the selection rule $\lambda^{\Delta Q}$, while the variance is dominated by cross-sector resonances at the edges; the variance law depends on whether the window fixes the number of states ($\sigma_G^2\propto\lambda$) or the energy interval ($\sigma_G^2\propto\lambda^2\ln\lambda^{-1}$), a size-bias effect that a resonance theory built from $\lambda=0$ data reproduces without free parameters. Once the golden-rule width exceeds the level spacing, the normalized variance becomes independent of the Hilbert-space dimension, and it reaches the exact random-matrix value $f(1-f)D/[(D-1)(D+2)]$ only when the width becomes comparable to the window, at breaking strengths of the order of the exchange coupling---long after the level statistics have become Wigner--Dyson. The random-walk estimate $e^{-S/2}$ is thus the correct endpoint, with a definite prefactor, of a crossover whose interior is controlled by edge resonances and by the ratio of the spreading width to the energy window.

\begin{acknowledgments}
I thank Dr.~Diptarka Das (IIT Kanpur) for valuable discussions.
\end{acknowledgments}

\section*{Data and code availability}
The code that generates every number and figure of this paper, together with the computed ensembles, is publicly available~\cite{code}.

% ======================================================================
\appendix
\onecolumngrid

\section{Proof of the edge representation (Theorem~\ref{thm:edge})}
\label{app:edge}

Let $H=H_0+\lambda V$ have eigenpairs $H|I\rangle=E_I|I\rangle$ with $E_I\neq E_J$ for $I\neq J$, and let $p_I=1$ if $I\in\mathcal W$ and $p_I=0$ otherwise, so that $\Pw=\sum_Ip_I|I\rangle\langle I|$.

\emph{Step 1: commutator with the charge.} Since $\hat Q|B\rangle=N_B|B\rangle$ and $\langle A|\hat Q=N_A\langle A|$,
\begin{align}
\langle A|[\hat Q,\Pw]|B\rangle&=\langle A|\hat Q\Pw|B\rangle-\langle A|\Pw\hat Q|B\rangle\nonumber\\
&=(N_A-N_B)\,G .
\label{eq:A1}
\end{align}

\emph{Step 2: matrix elements in the eigenbasis of $H$.} With $Q_{IJ}=\langle I|\hat Q|J\rangle$,
\begin{equation}
\langle I|[\hat Q,\Pw]|J\rangle=Q_{IJ}\,p_J-p_I\,Q_{IJ}=Q_{IJ}\,(p_J-p_I) .
\label{eq:A2}
\end{equation}
Only pairs with $p_I\neq p_J$ contribute; in particular $I\neq J$.

\emph{Step 3: off-diagonal elements of the charge.} Because $[H_0,\hat Q]=0$,
\begin{equation}
[H,\hat Q]=\lambda[V,\hat Q]=\lambda\hat C .
\label{eq:A3}
\end{equation}
Taking the $(I,J)$ matrix element, $\langle I|H\hat Q-\hat QH|J\rangle=(E_I-E_J)Q_{IJ}=\lambda C_{IJ}$, hence for $I\neq J$
\begin{equation}
Q_{IJ}=\frac{\lambda\,C_{IJ}}{E_I-E_J} .
\label{eq:A4}
\end{equation}
For $V=\sum_iS^x_i$ and $\hat Q=\sum_i(\tfrac12-S^z_i)$, $[S^x_i,\tfrac12-S^z_i]=-[S^x_i,S^z_i]=iS^y_i$, which gives Eq.~\eqref{eq:C}.

\emph{Step 4: assembly.} Inserting two resolutions of the identity into Eq.~\eqref{eq:A1} and using Eq.~\eqref{eq:A2},
\begin{align}
(N_A-N_B)\,G&=\sum_{I,J}\langle A|I\rangle\,Q_{IJ}(p_J-p_I)\,\langle J|B\rangle\nonumber\\
&=\sum_{I\notin\mathcal W}\sum_{J\in\mathcal W}\langle A|I\rangle Q_{IJ}\langle J|B\rangle-\sum_{I\in\mathcal W}\sum_{J\notin\mathcal W}\langle A|I\rangle Q_{IJ}\langle J|B\rangle .
\end{align}
Exchanging the names of the summation indices in the first sum, $I\leftrightarrow J$, both sums run over $I\in\mathcal W$, $J\notin\mathcal W$:
\begin{equation}
(N_A-N_B)\,G=\sum_{I\in\mathcal W}\sum_{J\notin\mathcal W}\Big[\langle A|J\rangle Q_{JI}\langle I|B\rangle-\langle A|I\rangle Q_{IJ}\langle J|B\rangle\Big].
\end{equation}
Dividing by $N_A-N_B\neq0$ and using Eq.~\eqref{eq:A4} for both $Q_{JI}$ and $Q_{IJ}$ gives Eq.~\eqref{eq:edge}. $\square$

\emph{Remarks.} (i)~For a real symmetric $H$ the eigenvectors are real, $\hat C$ is real antisymmetric, and $Q_{IJ}=Q_{JI}$ is real. (ii)~If $\Pw$ is replaced by $F(H)=\sum_IF(E_I)|I\rangle\langle I|$, Step 2 gives $Q_{IJ}[F(E_J)-F(E_I)]$ and Step 4 gives
\begin{equation}
\langle A|F(H)|B\rangle=\frac{\lambda}{N_A-N_B}\sum_{I,J}\langle A|I\rangle\,C_{IJ}\,\frac{F(E_J)-F(E_I)}{E_I-E_J}\,\langle J|B\rangle ,
\end{equation}
in which the divided difference is bounded by $\max|F'|$. Small denominators, and hence resonances, require a sharp window. (iii)~The proof uses no property of $\mathcal W$ other than being a set of eigenstates, so it applies to count and energy windows alike.

% ======================================================================
\section{Perturbative expansion of the window projector}
\label{app:dk}

\emph{Riesz projector.} For an energy window, let $\Gamma$ be a closed contour in the complex plane encircling the real interval $[E_-,E_+]$ and no other eigenvalue. Then $\Pw=(2\pi i)^{-1}\oint_\Gamma dz\,R(z)$ with $R(z)=(z-H)^{-1}$~\cite{Kato:1995}. With $R_0(z)=(z-H_0)^{-1}$, the second resolvent identity $R=R_0+\lambda R_0VR$ iterates to the Neumann series
\begin{equation}
R(z)=\sum_{k=0}^{\infty}\lambda^kR_0(z)\big[VR_0(z)\big]^k ,
\label{eq:B1}
\end{equation}
which converges on $\Gamma$ for $|\lambda|\,\|V\|<\min_{z\in\Gamma}\|R_0(z)\|^{-1}$.

\emph{Matrix elements.} In the charge-resolved eigenbasis of $H_0$, $R_0(z)=\sum_\alpha|\alpha\rangle\langle\alpha|/(z-e_\alpha)$, so
\begin{equation}
\langle\alpha|R_0(VR_0)^k|\beta\rangle=\sum_{\gamma_1,\dots,\gamma_{k-1}}\frac{V_{\alpha\gamma_1}V_{\gamma_1\gamma_2}\cdots V_{\gamma_{k-1}\beta}}{(z-e_\alpha)(z-e_{\gamma_1})\cdots(z-e_\beta)} .
\end{equation}

\emph{Contour integral and divided differences.} For distinct real nodes $x_0,\dots,x_k$, none on $\Gamma$, the residue theorem gives
\begin{equation}
\frac{1}{2\pi i}\oint_\Gamma\frac{dz}{\prod_{j=0}^k(z-x_j)}=\sum_{j:\,x_j\in\mathcal W}\ \prod_{i\neq j}\frac{1}{x_j-x_i}\equiv\chi_{\mathcal W}[x_0,\dots,x_k] ,
\label{eq:B3}
\end{equation}
which is the standard representation of the $k$th divided difference of the function equal to $1$ inside the contour and $0$ outside~\cite{Bhatia:1997}; for $k=1$, $\chi_{\mathcal W}[x_0,x_1]=[\chi_{\mathcal W}(x_0)-\chi_{\mathcal W}(x_1)]/(x_0-x_1)$. Inserting Eqs.~\eqref{eq:B1}--\eqref{eq:B3} into $G=\sum_{\alpha\beta}a_\alpha\langle\alpha|\Pw|\beta\rangle b_\beta$ gives Eq.~\eqref{eq:DK}. (Coincident nodes are handled by continuity; they occur with probability zero.)

\emph{Selection rule.} $V$ maps $\mathcal H_N$ into $\mathcal H_{N-1}\oplus\mathcal H_{N+1}$. A product $V_{\alpha\gamma_1}\cdots V_{\gamma_{k-1}\beta}$ with $|\alpha\rangle\in\mathcal H_{N_A}$ and $|\beta\rangle\in\mathcal H_{N_B}$ is nonzero only if the charge path has $k\ge|N_A-N_B|=\Delta Q$ steps and $k\equiv\Delta Q\pmod 2$. Hence $G^{(k)}=0$ for $k<\Delta Q$.

\emph{Edge localization at every order.} If all nodes lie inside $\mathcal W$, the sum in Eq.~\eqref{eq:B3} runs over all $j$ and equals the $k$th divided difference of the constant function $1$, which vanishes for $k\ge1$; if all lie outside, the sum is empty. Every nonzero term of Eq.~\eqref{eq:DK} therefore involves a chain of unperturbed energies that crosses a window edge.

\emph{Count windows.} For a count window with $L=(D-r)/2$, the edges lie in the gap between the $L$th and $(L+1)$th levels and in the gap between the $(L+r)$th and $(L+r+1)$th levels. As long as no level crosses an edge as $\lambda$ is switched on---i.e., away from resonances at the edges---the count window coincides with an energy window whose edges lie in those gaps, and the expansion applies unchanged. The resonances are treated non-perturbatively in Appendix~\ref{app:resonance}.

% ======================================================================
\section{Averages over Haar channels}
\label{app:channel}

Let $\bm a\in\mathbb R^{D_{N_A}}$ be uniformly distributed on the unit sphere of $\mathcal H_{N_A}$ and $\bm b$ independently on that of $\mathcal H_{N_B}$. By symmetry under $\bm a\to O\bm a$ for any orthogonal $O$, $\Expec[a_i]=0$ and $\Expec[a_ia_j]=c\,\delta_{ij}$, and $\sum_i\Expec[a_i^2]=1$ fixes $c=1/D_{N_A}$; likewise for $\bm b$. Writing $M=\hat\Pi_A\Pw\hat\Pi_B$ restricted to $\mathcal H_{N_B}\to\mathcal H_{N_A}$, $G=\bm a^{\!\top}M\bm b$, so $\Expec[G]=0$ and
\begin{align}
\Expec[G^2]&=\Expec\big[\bm a^{\!\top}M\bm b\,\bm b^{\!\top}M^{\!\top}\bm a\big]
=\Expec_{\bm a}\Big[\bm a^{\!\top}M\,\frac{\mathbb 1}{D_{N_B}}\,M^{\!\top}\bm a\Big]\nonumber\\
&=\frac{\Tr(MM^{\!\top})}{D_{N_A}D_{N_B}}=\frac{\Tr(\hat\Pi_A\Pw\hat\Pi_B\Pw)}{D_{N_A}D_{N_B}} ,
\end{align}
which is Eq.~\eqref{eq:T}. In the eigenbasis of $H$, $\Tr(\hat\Pi_A\Pw\hat\Pi_B\Pw)=\sum_{I,J\in\mathcal W}(\hat\Pi_A)_{IJ}(\hat\Pi_B)_{JI}$. Since $\hat\Pi_A\hat\Pi_B=0$, $\sum_{J}(\hat\Pi_A)_{IJ}(\hat\Pi_B)_{JI}=0$ for every $I$, so the sum over $J\in\mathcal W$ equals minus the sum over $J\notin\mathcal W$, which gives Eq.~\eqref{eq:Tedge}. Numerically we evaluate Eq.~\eqref{eq:T} as $\|P_{AB}\|_F^2/(D_{N_A}D_{N_B})$, where $P_{AB}$ is the $(A,B)$ block of $\Pw$ in the computational basis.

% ======================================================================
\section{Proofs of the random-projector theorem}
\label{app:rmt}

\emph{Proof by invariance (orthogonal class).} Let $\hat P$ be a Haar-random rank-$r$ real projector. Its distribution is invariant under $\hat P\to O\hat PO^{\!\top}$, so the fourth-rank tensor $\Expec[P_{ab}P_{cd}]$ is an invariant of the orthogonal group and symmetric under $a\leftrightarrow b$ and $c\leftrightarrow d$. The only such tensors are built from Kronecker deltas,
\begin{equation}
\Expec[P_{ab}P_{cd}]=\alpha\,\delta_{ab}\delta_{cd}+\beta\,(\delta_{ac}\delta_{bd}+\delta_{ad}\delta_{bc}) .
\end{equation}
Hence $\Expec[P_{12}^2]=\beta$, $\Expec[P_{11}P_{22}]=\alpha$, and $\Expec[P_{11}^2]=\alpha+2\beta$. Two exact constraints fix $\alpha$ and $\beta$. From $\hat P^2=\hat P$, $\sum_bP_{1b}P_{b1}=P_{11}$, whose expectation is $r/D$:
\begin{equation}
(\alpha+2\beta)+(D-1)\beta=\frac rD .
\end{equation}
From $(\Tr\hat P)^2=r^2$, $\sum_{a,c}\Expec[P_{aa}P_{cc}]=r^2$:
\begin{equation}
D(\alpha+2\beta)+D(D-1)\alpha=D^2\alpha+2D\beta=r^2 .
\end{equation}
Eliminating $\alpha=r/D-(D+1)\beta$ from the first equation,
\begin{align}
D^2\Big[\frac rD-(D+1)\beta\Big]+2D\beta&=r^2\nonumber\\
\Rightarrow\quad \beta\,D(D^2+D-2)&=rD-r^2 ,
\end{align}
so $\beta=r(D-r)/[D(D-1)(D+2)]$. With $r=fD$ this is $\sigCOE^2$ of Eq.~\eqref{eq:rmt}. For fixed orthonormal $|a\rangle,|b\rangle$, rotate them to the first two basis vectors.

\emph{Unitary class.} Invariance under $\hat P\to U\hat PU^\dagger$ allows $\Expec[P_{ab}P_{cd}]=\alpha\,\delta_{ab}\delta_{cd}+\beta\,\delta_{ad}\delta_{bc}$, so $\Expec|P_{12}|^2=\Expec[P_{12}P_{21}]=\beta$ and $\Expec[P_{11}^2]=\alpha+\beta$. The constraints become $\alpha+D\beta=r/D$ and $D^2\alpha+D\beta=r^2$, giving $\beta=r(D-r)/[D(D^2-1)]=f(1-f)D/(D^2-1)$.

\emph{Proof via the Beta law.} Let $w=P_{11}$, the squared norm of the projection of a fixed unit vector onto the random subspace. For a real (complex) Haar subspace this equals the squared norm of the first $r$ components of a random unit vector in $\mathbb R^D$ ($\mathbb C^D$), so $w\sim\mathrm{Beta}(\tfrac r2,\tfrac{D-r}2)$ [$\mathrm{Beta}(r,D-r)$]~\cite{Zyczkowski:2001}, with
\begin{equation}
\Var(w)=\frac{2r(D-r)}{D^2(D+2)}\quad\Big[\frac{r(D-r)}{D^2(D+1)}\Big] .
\end{equation}
All off-diagonal elements have the same distribution, and $\sum_{a\neq b}|P_{ab}|^2=\Tr\hat P^2-\sum_aP_{aa}^2=r-\sum_aP_{aa}^2$. Taking expectations,
\begin{equation}
D(D-1)\,\Expec|P_{12}|^2=r-D\Big[\frac{r^2}{D^2}+\Var(w)\Big] ,
\end{equation}
and inserting $\Var(w)$ reproduces Eq.~\eqref{eq:rmt} in both classes. $\square$ The same result follows from the Weingarten calculus for Haar integrals~\cite{Collins:2006}.

\emph{Asymptotics.} $D/[(D-1)(D+2)]=D^{-1}[1-D^{-1}+O(D^{-2})]$ and $D/(D^2-1)=D^{-1}[1+O(D^{-2})]$, giving Eq.~\eqref{eq:rmtasym}.

\begin{table}[t]
\caption{\label{tab:mc}Monte Carlo check of Theorem~\ref{thm:rmt}: ratio of the sampled $\Expec|P_{12}|^2$ to Eq.~\eqref{eq:rmt}, with one-standard-error uncertainty ($10^5$ samples for $D\le64$, $2\times10^4$ for $D=256$).}
\begin{ruledtabular}
\begin{tabular}{ccccc}
$D$ & $f$ & orthogonal & unitary & $\sigma^2_{\rm RMT}$ (orth.)\\
\colrule
16 & 0.25 & $0.9903(45)$ & $0.9970(32)$ & $1.111\times10^{-2}$\\
16 & 0.5 & $0.9993(41)$ & $1.0030(30)$ & $1.481\times10^{-2}$\\
64 & 0.25 & $1.0042(45)$ & $0.9956(32)$ & $2.886\times10^{-3}$\\
64 & 0.5 & $1.0017(44)$ & $1.0059(31)$ & $3.848\times10^{-3}$\\
256 & 0.25 & $1.0062(101)$ & $1.0052(72)$ & $7.296\times10^{-4}$\\
256 & 0.5 & $0.9979(98)$ & $1.0044(70)$ & $9.728\times10^{-4}$\\
\end{tabular}
\end{ruledtabular}
\end{table}

% ======================================================================
\section{Resonance statistics at the window edges}
\label{app:resonance}

\emph{Two-level reduction.} Consider $|\alpha\rangle\in\mathcal H_{N_A}$ and $|\beta\rangle\in\mathcal H_{N_A+1}$ with $s=e_\alpha-e_\beta$ and coupling $g=\lambda V_{\alpha\beta}$, other levels being far away on the scale of $|s|$ and $|g|$. The block $h_2=\tfrac12(e_\alpha+e_\beta)\mathbb 1+\tfrac12 s\,\sigma^z+g\,\sigma^x$ has eigenvalues $\tfrac12(e_\alpha+e_\beta)\pm\tfrac12\sqrt{s^2+4g^2}$ and eigenvectors $|+\rangle=\cos\theta|\alpha\rangle+\sin\theta|\beta\rangle$, $|-\rangle=-\sin\theta|\alpha\rangle+\cos\theta|\beta\rangle$, with $\tan2\theta=2g/s$. If exactly one of them lies in $\mathcal W$,
\begin{equation}
\langle\alpha|\Pw|\beta\rangle=\pm\cos\theta\sin\theta=\pm\tfrac12\sin2\theta=\pm\frac{g}{\sqrt{s^2+4g^2}} ,
\end{equation}
which is Eq.~\eqref{eq:x}. If both lie in $\mathcal W$, $\langle\alpha|\Pw|\beta\rangle=\cos\theta\sin\theta-\sin\theta\cos\theta=0$. For $|s|\gg|g|$, $x\to g/s$, the first-order term of Eq.~\eqref{eq:G1}. The channel average of the pair's contribution is $\Expec[a_\alpha^2]\Expec[b_\beta^2]x^2=x^2/(D_{N_A}D_{N_B})$.

\emph{Pair densities.} At $\lambda=0$ the spectra of different sectors are statistically independent. The number of pairs $(\alpha,\beta)\in\mathcal H_{N_A}\times\mathcal H_{N_B}$ with midpoint in $[m,m+dm]$ and splitting $s=e_\alpha-e_\beta\in[s,s+ds]$ is therefore $\rho_A(m)\rho_B(m)\,dm\,ds$ for $|s|$ small compared with the scale on which the densities vary, with $s\in\mathbb R$ covering both orderings.

\emph{Count window.} The upper edge lies in the gap between consecutive levels $k$ and $k+1$ of the full spectrum, with $k$ fixed. Over the ensemble, this gap is a uniformly chosen gap near the energy $E_+$. The number of gaps per unit energy is $\rho$. The number of $(A,B)$ pairs with $|s|\in[s,s+ds]$ per unit energy is $2\rho_A\rho_B\,ds$, and for $s\to0$ the probability that a third level lies between them vanishes, so they are adjacent. The probability density of the selected gap being an $(A,B)$ gap of splitting $|s|$ is thus $2\rho_A\rho_B/\rho$ as $s\to0$; summing the two edges gives Eq.~\eqref{eq:nu}. The resonant contribution is
\begin{equation}
\sigma_{G,\rm res}^2=\frac{\nu}{D_{N_A}D_{N_B}}\int_0^\infty ds\,\overline{\frac{g^2}{s^2+4g^2}}=\frac{\nu}{D_{N_A}D_{N_B}}\,\overline{|g|}\,\frac{\pi}{4} ,
\end{equation}
using $\int_0^\infty ds\,g^2/(s^2+4g^2)=g^2\cdot\pi/(2\cdot2|g|)$. This is Eq.~\eqref{eq:count}. Extending the flat density to $s=\infty$ overcounts by $O(\nu g^2/\Delta)$, an $O(\lambda^2)$ term of the same order as the non-resonant pairs.

\emph{Distribution of the resonant contribution.} For a single edge pair, $x^2>y$ (with $0<y<\tfrac14$) if and only if $s^2<g^2(y^{-1}-4)$. With the flat density $\nu$, $P(x^2>y)=\nu\,\overline{|g|}\sqrt{y^{-1}-4}$, which is Eq.~\eqref{eq:tail}. The cutoff at $y=\tfrac14$ is the resonant maximum $|x|=\tfrac12$.

\emph{Energy window.} For a fixed energy $E_+$, a pair with midpoint $m$ has hybridized levels $m\pm\tfrac12\sqrt{s^2+4g^2}$, and exactly one lies below $E_+$ if and only if $|m-E_+|<\tfrac12\sqrt{s^2+4g^2}$. Integrating over $m$ gives the measure $\sqrt{s^2+4g^2}$, so the contribution of edge $E_+$ is
\begin{equation}
\frac{\rho_A\rho_B}{D_{N_A}D_{N_B}}\int_{-s_c}^{s_c}ds\,\overline{\sqrt{s^2+4g^2}\,\frac{g^2}{s^2+4g^2}}=\frac{2\rho_A\rho_B}{D_{N_A}D_{N_B}}\,\overline{g^2\,\mathrm{arsinh}\frac{s_c}{2|g|}} ,
\end{equation}
which grows as $2\rho_A\rho_B\overline{g^2}\ln(s_c/|g|)$; the cutoff $s_c\sim\omega_c$ is set by the decay of $\overline{|V_{\alpha\beta}|^2}$ with $|e_\alpha-e_\beta|$. This gives Eq.~\eqref{eq:energy}. The small-$s$ region contributes only an $O(g^2)$ amount without the logarithm, in contrast to the count window. Geometrically, a fixed energy falls into a gap with probability proportional to the gap length (length-biased sampling~\cite{Feller:1971}), so near-degenerate gaps are rarely selected [Fig.~\ref{fig:setup}(b)].

\begin{figure}[t]
\centering
\includegraphics[width=0.5\textwidth]{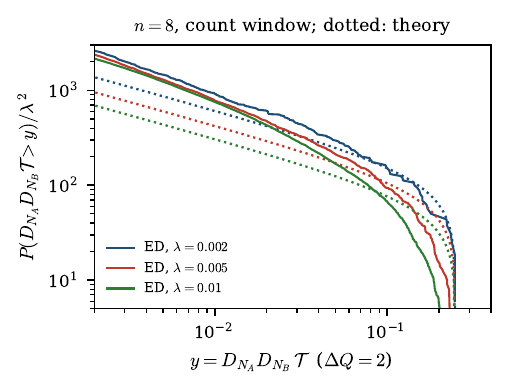}
\caption{\label{fig:dq2}Resonant tail for $\Delta Q=2$ ($n=8$, count window, 40000 realizations per $\lambda$), rescaled by $\lambda^2$. Solid: exact diagonalization; dotted: $\nu_2\overline{|g_{\rm eff}|}\lambda^{-2}\sqrt{y^{-1}-4}$ from $\lambda=0$ data.}
\end{figure}

\emph{$\Delta Q=2$.} For $|\alpha\rangle\in\mathcal H_{N_A}$ and $|\beta\rangle\in\mathcal H_{N_A+2}$ nearly degenerate, eliminating the intermediate sector to second order (Schrieffer--Wolff) gives the effective coupling
\begin{equation}
g_{\rm eff}=\frac{\lambda^2}{2}\sum_{\gamma\in\mathcal H_{N_A+1}}V_{\alpha\gamma}V_{\gamma\beta}\Big(\frac{1}{e_\alpha-e_\gamma}+\frac{1}{e_\beta-e_\gamma}\Big) ,
\end{equation}
and the arguments above apply with $g\to g_{\rm eff}$ and with the density $\nu_2=\sum_e2\rho_A\rho_{N_A+2}/\rho$ of edge-adjacent $(N_A,N_A+2)$ gaps. The distribution of $g_{\rm eff}$ is itself broad because of near-resonant intermediate states; we regularize each denominator at the scale of its own coupling, $1/d\to d/(d^2+\lambda^2V_{\alpha\gamma}^2+\lambda^2V_{\gamma\beta}^2)$, which makes $\overline{|g_{\rm eff}|}/\lambda^2$ weakly (logarithmically) $\lambda$ dependent. Figure~\ref{fig:dq2} compares the resulting parameter-free tail with exact diagonalization.

\emph{Evaluation from $\lambda=0$ data.} For each $\lambda=0$ realization we compute the charge-resolved spectra and eigenvectors, the matrix $V_{\alpha\beta}$ between sectors $N_A$ and $N_A+1$, the local densities of states at the edges, and the golden-rule coefficients $\gamma_\alpha$ of Eq.~\eqref{eq:Gamma} (with the $\delta$ function broadened to a Gaussian of width $5\Delta$). Pairs with $|s|\ge s_0=\Delta/4$ are summed exactly with weight $x^2=g^2/(s^2+4g^2+\Gamma_{\alpha\beta}^2)$, $\Gamma_{\alpha\beta}=\tfrac12(\gamma_\alpha+\gamma_\beta)\lambda^2$; for the energy window a pair is counted if exactly one of its hybridized levels lies in $[E_-,E_+]$. Pairs with $|s|<s_0$ are integrated analytically with the densities above,
\begin{align}
\text{count:}\quad&\nu\ \overline{\frac{g^2}{a}\arctan\frac{s_0}{a}},\qquad a=\sqrt{4g^2+\Gamma^2},\\
\text{energy:}\quad&\sum_{e=\pm}\rho_A\rho_B\ \overline{\int_{-s_0}^{s_0}\!ds\,\frac{g^2\sqrt{s^2+4g^2}}{s^2+4g^2+\Gamma^2}} ,
\end{align}
where the bar averages over the empirical distribution of $|V_{\alpha\beta}|$ and $\gamma_{\alpha\beta}$ of near-degenerate cross-sector pairs at the edges. The ensemble sizes are $6000$ ($n=8$), $3000$ ($n=10$), and $400$ ($n=12$) realizations.

% ======================================================================
\section{Golden-rule regime}
\label{app:golden}

For $\Gamma\gg\Delta$ an unperturbed state is spread over a Breit--Wigner distribution of width $\Gamma$ and the response of the window projector to a pair at energy separation $s$ is cut off at $|s|\sim\Gamma$; we model this by $x^2=g^2/(s^2+\Gamma^2)$ for $|g|\ll\Gamma$. A pair straddles an edge when the edge lies between its two energies, which has measure $|s|$ in the midpoint. In the continuum limit, for each edge,
\begin{equation}
\sigma_G^2\Big|_{e}\simeq\frac{\rho_A\rho_B}{D_{N_A}D_{N_B}}\int_{-\omega_c}^{\omega_c}ds\,|s|\,\frac{\lambda^2\overline{V^2}}{s^2+\Gamma^2}
=\frac{\rho_A\rho_B\lambda^2\overline{V^2}}{D_{N_A}D_{N_B}}\ln\Big(1+\frac{\omega_c^2}{\Gamma^2}\Big).
\end{equation}
Fermi's golden rule for the partial width into sector $N_B$, $\Gamma_{A\to B}=2\pi\lambda^2\overline{V^2}\rho_B$, eliminates $\overline{V^2}$:
\begin{equation}
\sigma_G^2\simeq\frac{1}{\pi D_{N_A}D_{N_B}}\sum_{e=\pm}\rho_A(E_e)\,\Gamma_{A\to B}\ln\frac{\omega_c}{\Gamma} .
\end{equation}
Dividing by $\sigCOE^2\simeq f(1-f)/D$ and writing $\rho_A=D_{N_A}\varrho_A$ gives Eq.~\eqref{eq:golden}. The normalized density $\varrho_A$ is of order the inverse spectral width ($\propto n^{-1/2}$), $\Gamma_{A\to B}\propto n\lambda^2$ since $\sum_\beta|V_{\alpha\beta}|^2=\langle\alpha|V^2|\alpha\rangle=O(n)$ (note $\langle\alpha|V|\alpha\rangle=0$), and $D/D_{N_B}=2^n/\binom{n}{n/2+1}\propto\sqrt{n}$, so the right-hand side is independent of $D$ up to factors polynomial in $n$. The same estimate shows that $\sigma_G^2/\sigCOE^2$ becomes of order one when $\Gamma$ is of order the window width, which defines $\lB$ in Eq.~\eqref{eq:lD}.

For comparison, in the resonant regime $\sigma_G^2/\sigCOE^2=\tilde c_1\lambda$ with $\tilde c_1=(\pi/4)\nu\overline{|V|}D/[f(1-f)D_{N_A}D_{N_B}]$. Eigenstate thermalization implies $\overline{V^2}\simeq\Gamma_{A\to B}/(2\pi\lambda^2\rho_B)\propto\rho_B^{-1}$, so $\overline{|V|}\propto D^{-1/2}$, while $\nu D/(D_{N_A}D_{N_B})$ is intensive. Hence $\tilde c_1\propto D^{-1/2}$, and the two expressions match at $\lambda\sim\lD$, where $\Gamma\sim\Delta$.

% ======================================================================
\section{Numerical methods and checks}
\label{app:numerics}

\emph{Construction and diagonalization.} $H(\lambda)$ is built as a dense real symmetric $2^n\times2^n$ matrix in the computational basis and diagonalized with LAPACK (\texttt{numpy.linalg.eigh}). For the $\lambda=0$ resonance theory we diagonalize each charge sector separately. Every script uses a seeded pseudorandom generator; all numbers in this paper are regenerated by the code of Ref.~\cite{code}.

\emph{Ensembles.} For $n=8$: $4000$ realizations per $\lambda\le3\times10^{-3}$ and $2000$ otherwise; $n=10$: $1500$ and $300$; $n=12$: $30$ per $\lambda$ ($\lambda\ge10^{-2}$). At $\lambda=0$, $20$ ($n\le10$) and $4$ ($n=12$) realizations suffice because $G=0$ identically. For each realization and $\lambda$ we store $\mathcal T$ [Eq.~\eqref{eq:T}] for both windows and $\Delta Q=1,2$, two sampled values of $G$ per $\Delta Q$ and window (fresh Haar channels), the rank of the energy window, and $\langle r\rangle$ of the central $80\%$ of the full spectrum. For the energy window, $\mathcal T$ is normalized realization by realization by $\sigma_{\rm RMT}^2(D,r_{\mathcal W}/D)$ with the actual rank $r_{\mathcal W}$. Error bars are bootstrap standard errors over realizations.

\begin{figure}[t]
\centering
\includegraphics[width=0.5\textwidth]{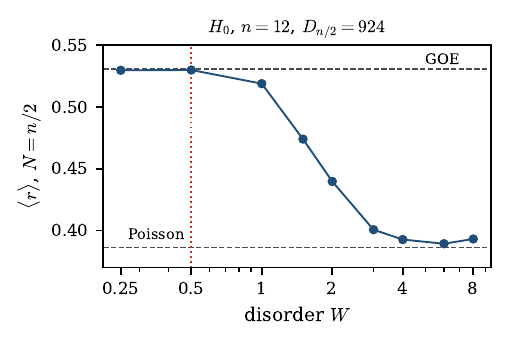}
\caption{\label{fig:calib}Level-spacing ratio of $H_0$ in the half-filling sector ($n=12$, $60$ realizations per point). The dotted line marks $W=0.5$ used throughout.}
\end{figure}

\emph{Calibration.} Figure~\ref{fig:calib} shows $\langle r\rangle$ of $H_0$ in the half-filling sector versus $W$ for $n=12$; at $W=0.5$, $\langle r\rangle=0.530(2)$.

\emph{Exact identities.} Over random realizations with $n\in\{6,8,10\}$ and $\lambda\in\{10^{-3},10^{-1},1\}$, the edge representation~\eqref{eq:edge} agrees with direct evaluation to a maximal relative error of $7.0\times10^{-11}$, the edge-trace form~\eqref{eq:Tedge} to $2.0\times10^{-14}$, and Monte Carlo averages over $3000$ Haar channel pairs agree with Eq.~\eqref{eq:T} within their statistical error (ratio $1.007\pm0.030$). At $\lambda=0$, $\max|G|=4.9\times10^{-16}$. The first-order formula~\eqref{eq:G1} (energy window, $n=8$) agrees with exact diagonalization to relative error $5.8\times10^{-8}$, $1.1\times10^{-8}$, $1.6\times10^{-6}$ at $\lambda=10^{-7},10^{-6},10^{-5}$, consistent with $O(\lambda)$ corrections.

\emph{Pair theory on identical realizations.} For the same disorder realizations, we compare the exact $\mathcal T$ of $H(\lambda)$ with the pair sum $\sum x^2/(D_{N_A}D_{N_B})$ over straddling $(A,B)$ pairs evaluated from the $\lambda=0$ eigenpairs. At $\lambda=10^{-4}$ the median relative deviation per realization is $4.0\times10^{-6}$ (count) and $1.5\times10^{-6}$ (energy) for $n=8$, and $2.2\times10^{-5}$ and $7.4\times10^{-6}$ for $n=10$; at $\lambda=10^{-3}$ it is $3.1\times10^{-4}$ and $1.6\times10^{-4}$ ($n=8$). At $\lambda=10^{-2}$, where $\Gamma/\Delta=O(1)$, the unbroadened pair sum overestimates the mean by $17\%$ (count) and $13\%$ (energy) for $n=8$, which is the overlap of resonances that the golden-rule broadening of Sec.~\ref{sec:crossover} accounts for. The agreement on identical realizations at small $\lambda$ shows that the residual differences between theory and ensemble averages in Fig.~\ref{fig:resonance}(a) at the smallest $\lambda$ come from the sampling of rare resonances, not from the theory.

\emph{Fixed channels.} If the channels are fixed and only the disorder is averaged, $\sigma_G$ depends on the channel. For six fixed Haar pairs at $n=8$ and $\lambda=0.7$ we find $\sigma_G/\sigma_{\rm RMT}=0.49, 0.52, 0.62, 0.76, 0.67, 1.21$, against $1.05$ for the exact channel average. Each channel has a definite energy distribution with respect to $H_0$, which the disorder does not randomize; its overlap with the energy shell therefore biases its own variance. Estimates from a few fixed channel pairs are not reliable, which is why we use Eq.~\eqref{eq:T}.

% ======================================================================
\section{Gate-level implementation}
\label{app:circuit}

\emph{Pauli form and Trotter circuit.} In Pauli operators,
\begin{equation}
H(\lambda)=\sum_{i=1}^{n-1}\frac{X_iX_{i+1}+Y_iY_{i+1}+Z_iZ_{i+1}}{4}+\sum_{i=1}^n\frac{h_i}{2}Z_i+\frac{\lambda}{2}\sum_{i=1}^nX_i .
\label{eq:pauli}
\end{equation}
We compile $e^{-iHt}$ with the second-order Suzuki--Trotter product~\cite{Lloyd:1996,Suzuki:1991,Childs:2021},
\begin{equation}
e^{-iHt}\simeq\Big[\prod_{k=1}^{K}e^{-iH_k\delta t/2}\prod_{k=K}^{1}e^{-iH_k\delta t/2}\Big]^{t/\delta t},
\end{equation}
where $H_k$ are the $K=5n-3$ Pauli terms of Eq.~\eqref{eq:pauli}, ordered bond by bond. A two-qubit rotation $e^{-i\theta P_iP_j}$ with $P\in\{X,Y,Z\}$ is a CNOT--$R_z(2\theta)$--CNOT ladder conjugated by single-qubit Clifford gates that map $P$ to $Z$; single-qubit terms are $R_z$ and $R_x$ rotations. Since $[X_iX_j,Y_iY_j]=[X_iX_j,Z_iZ_j]=0$, the three rotations of one bond multiply to the exact bond propagator, which conserves $\hat Q$; hence at $\lambda=0$ every Trotter step conserves the charge exactly, and any measured leakage between charge sectors on hardware is an error event. In Qiskit~\cite{Qiskit:2024} the product is a \texttt{PauliEvolutionGate} with \texttt{SuzukiTrotter(order=2)} synthesis.

\begin{figure*}[t]
\centering
\includegraphics[width=0.85\textwidth]{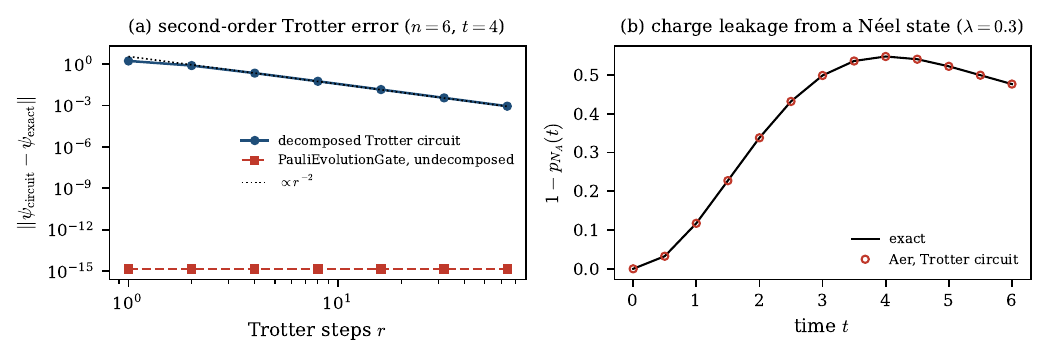}
\caption{\label{fig:qiskit}Qiskit validation of the time-evolution circuit. (a)~Error of the second-order Trotter circuit versus the number of steps (circles) and the spurious machine-precision agreement of an undecomposed \texttt{PauliEvolutionGate} (squares). (b)~Charge leakage from a N\'eel state under $H(\lambda=0.3)$: exact evolution (line) and Aer simulation of the decomposed Trotter circuit (symbols).}
\end{figure*}

\emph{A simulation pitfall.} \texttt{Statevector} simulation of a circuit containing an undecomposed \texttt{PauliEvolutionGate} evaluates the gate through its exact matrix exponential, irrespective of the requested synthesis: the result agrees with $e^{-iHt}$ to machine precision for any number of Trotter steps and hides the Trotter error entirely [Fig.~\ref{fig:qiskit}(a)]. After explicit decomposition, the error decreases as $r^{{-1.99}}$ with the number of steps $r$, as expected for a second-order formula. All circuit results below are obtained from decomposed circuits.

\emph{Validation.} Starting from a N\'eel state in the half-filling sector ($n=6$, $\lambda=0.3$), we measure the probability $1-p_{N_A}(t)$ of having left the initial charge sector, which is diagonal in the computational basis. With $8$ Trotter steps per unit time, the Aer statevector simulation of the decomposed circuit reproduces exact evolution with a maximal deviation of $9.3\times10^{-5}$ in any charge-sector population for $t\le6$ [Fig.~\ref{fig:qiskit}(b)]; at $\lambda=0$ the circuit leakage is $4.5\times10^{-13}$, at the level of floating-point rounding.

\emph{Resources.} After transpilation to the basis $\{\mathrm{CX},R_z,\sqrt X,X\}$ on a linear coupling map (Qiskit optimization level 1), one second-order step costs 84, 108, 132, 180, 228 CNOT gates for $n=8,10,12,16,20$. A dedicated decomposition of each Heisenberg bond into three CNOTs would reduce this to $3(n-1)$ per first-order step.

\twocolumngrid
\bibliography{refs}

\end{document}